\documentclass[]{tPHM2e}

\begin{document}
\doi{10.1080/1478643YYxxxxxxxx}
\issn{1478-6443}
\issnp{1478-6435}
\jvol{00} \jnum{00} \jyear{2009} \jmonth{}

\markboth{S.L. Bud'ko, N. Ni, and P.C. Canfield}{Philosophical Magazine}

\articletype{}

\title{Anisotropic thermal expansion of {\it AE}Fe$_2$As$_2$ ({\it AE} = Ba, Sr, Ca) single crystals.}

\author{Sergey L. Bud'ko, Ni Ni, and Paul C. Canfield\\\vspace{6pt}  {\em{Ames Laboratory U.S. DOE and Department of Physics and Astronomy, Iowa State University, Ames, Iowa 50011, USA}}\\\vspace{6pt}\received{ } }

\maketitle

\begin{abstract}

We report anisotropic thermal expansion of the parent, {\it AE}Fe$_2$As$_2$ ({\it AE} = Ba, Sr, and Ca), compounds. Above the structural/antiferromagnetic phase transition  anisotropy of the thermal expansion coefficients is observed, with the coefficient along the $a$-axis being significantly smaller than the coefficient for the $c$-axis. The high temperature ($200$ K $\leq T \leq 300$ K) coefficients themselves have similar values for the compounds studied. The sharp anomalies associated with the structural/antiferromagnetic phase transitions are clearly seen in the thermal expansion measurements. For all three pure compounds the "average" $a$-value increases and the $c$-lattice parameter decreases on warming through the transition with the smallest change in the lattice parameters observed for SrFe$_2$As$_2$. The data are in general agreement with the literature data from X-ray and neutron diffraction experiments.

\bigskip

\end{abstract}

For over a year many experimental and theoretical aspects of physics of ternary {\it AE}{\it TM}$_2$As$_2$ ({\it AE} = Ba, Sr, and Ca; {\it TM} = transition metal) compounds were studied in great detail. One of the motivations for such studies was a desire to shed some light on what are the salient parameters that govern doping- or pressure-induced superconductivity \cite{rot08a,sef08a,tor08a,yuw08a,ali09a,can09b,nin09a} in these materials.  The common feature of the parent compounds \cite{rot08b,hua08a,yan08a,jes08a,teg08a,nin08a,gol08a} is a distinct, moderately high temperature (above $\sim 100$ K), coupled, structural/antiferromagnetic phase transition that generally is argued to be of the first order. Thermal expansion is often claimed to be uniquely sensitive to just such magnetic, structural and superconducting transitions \cite{kro98a}. For example, in Co-doped BaFe$_2$As$_2$, anisotropic thermal expansion measurements \cite{bud09a,har09a,luz09a} have been instrumental in inferring unusually large, anisotropic, uniaxial pressure derivatives of superconducting transition temperature. For the parent {\it AE}{\it TM}$_2$As$_2$ materials though, we are aware only of data from pure BeFe$_2$As$_2$ \cite{bud09a}. In this work we present and compare anisotropic thermal expansion measurements for three parent compounds, {\it AE}Fe$_2$As$_2$ ({\it AE} = Ba, Sr, and Ca). (Data for BaFe$_2$As$_2$ are taken from \cite{bud09a}.) For completeness, data on Sn - grown BeFe$_2$As$_2$ \cite{nin08b} are included as well.
\\

Single crystals of {\it AE}Fe$_2$As$_2$ ({\it AE} = Ba, Sr, and Ca) were grown using conventional high temperature solution growth technique \cite{can92a}. Details of growth conditions and physical properties of these samples are outlined in \cite{yan08a,nin08a,nin08b,nin08c}. It should be noted, though, that BaFe$_2$As$_2$ samples grown out of FeAs as well as Sn have been examined. The SrFe$_2$As$_2$ and CaFe$_2$As$_2$ samples were both grown out of Sn. The BaFe$_2$As$_2$/FeAs, SrFe$_2$As$_2$/Sn and CaFe$_2$As$_2$/Sn samples are considered to be pure and well ordered whereas BaFe$_2$As$_2$/Sn is known to be slightly Sn-doped. Thermal expansion data were obtained using a capacitive dilatometer constructed of OFHC copper, mounted in a Quantum Design PPMS instrument. A detailed description of the dilatometer is presented elsewhere \cite{sch06a}. The samples were cut and lightly polished so as to have parallel surfaces approximately parallel to the $a$ in-plane direction and parallel to the $c$-direction with the distances $L$ between the surfaces ranging between approximately $0.3 - 3$ mm. Measurements were performed on warming. Volume thermal expansivity, $\Delta V/V_0$, and thermal expansion coefficient, $\beta$ are calculated as $\Delta V/V_0 = 2 \cdot \Delta a/a_0 + \Delta c/c_0$ and $\beta = 2 \cdot \alpha_a + \alpha_c$ respectively. We need to note that (i) the {\it AE}Fe$_2$As$_2$ crystals are rather soft and micaceous and require more than usual care in shaping; this makes the results more prone to errors due the morphology of the samples; (ii) below the structural/antiferromagnetic transition the samples are comprised of structural domains \cite{tan09a} so that low temperature, in-plane, bulk thermal expansion measurements present some average of two in-plane directions.
\\

\begin{figure}
\begin{center}
\begin{minipage}{130mm}
\subfigure[]{
\resizebox*{6.5cm}{!}{\includegraphics{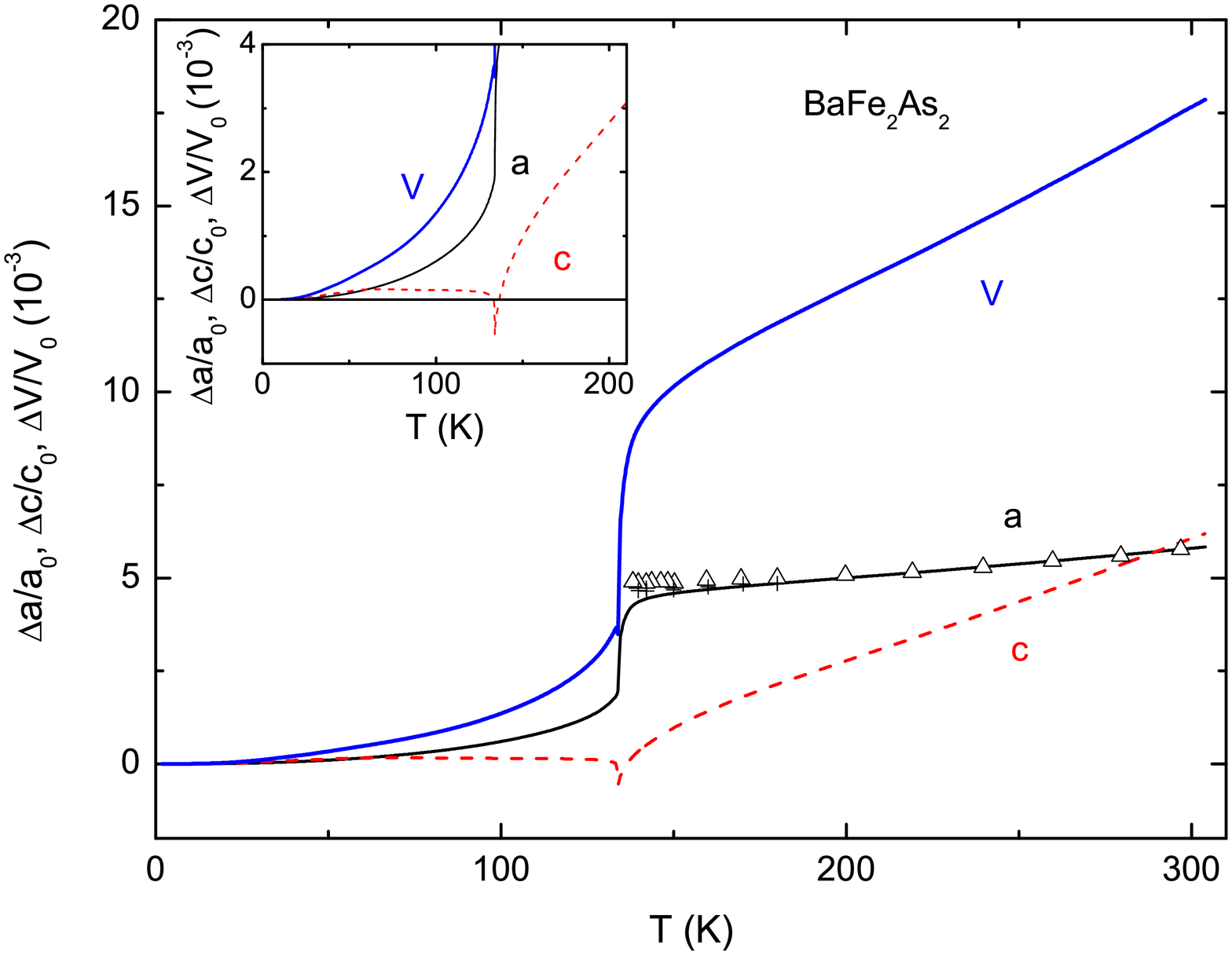}}}%
\subfigure[]{
\resizebox*{6.5cm}{!}{\includegraphics{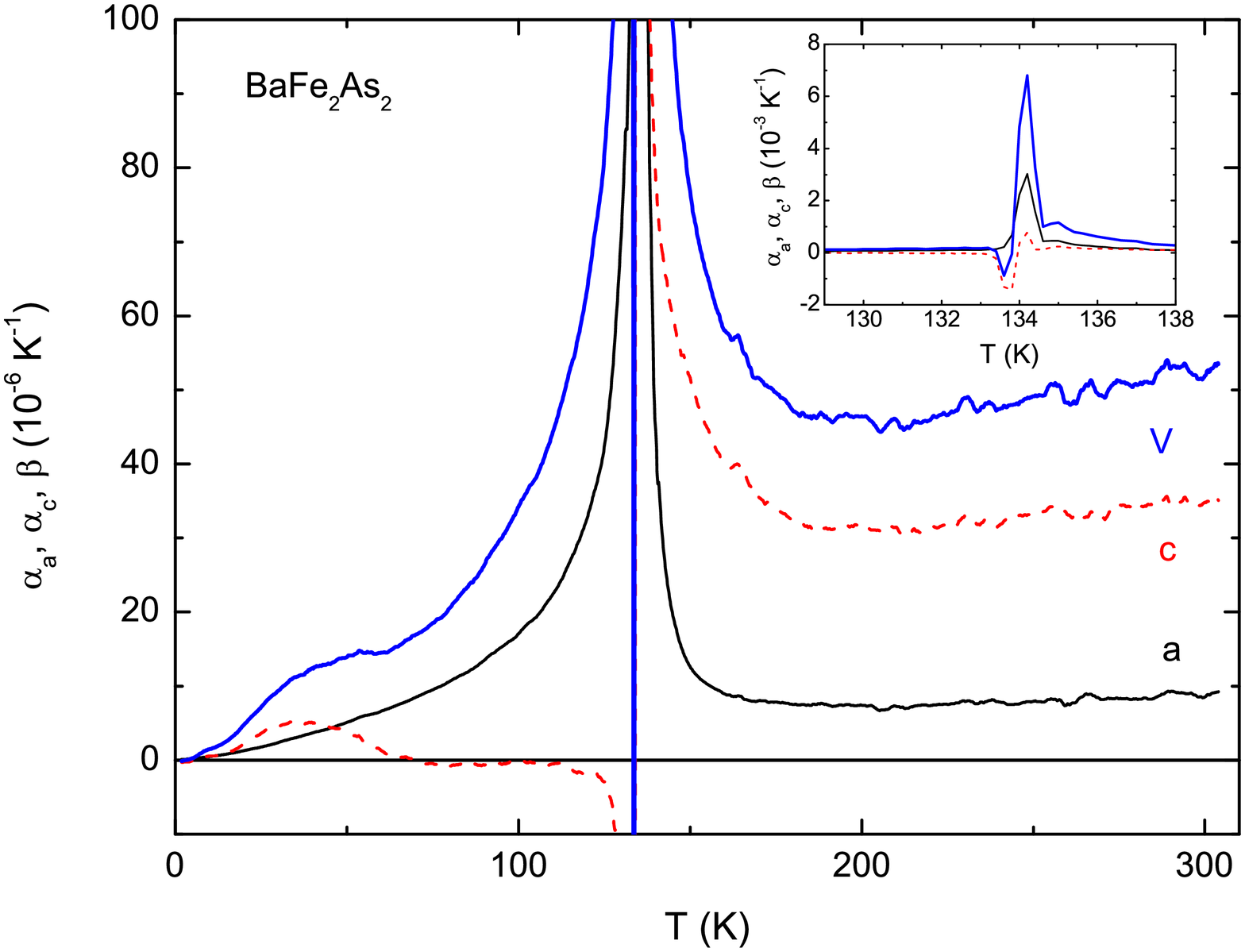}}}%
\caption{(a) Anisotropic thermal expansivities of BaFe$_2$As$_2$ plotted relative to the values at $T = 1.8$ K; inset: enlarged low temperature part. Symbols: $a$-axis data from \cite{rot09a} (triangles, normalized at 298 K) and from \cite{rot08b} (crosses, normalized at 180 K) (b) Anisotropic thermal expansion coefficients of BaFe$_2$As$_2$; inset: enlarged region near structiral/magnetic phase transitions. Data from Ref. \cite{bud09a}.}%
\label{F1}
\end{minipage}
\end{center}
\end{figure}

Anisotropic thermal expansivities (relative to the values at 1.8 K) and thermal expansion coefficients for pure BaFe$_2$As$_2$ are shown in Fig. \ref{F1}. Features corresponding to the structural/antiferromagnetic transition are clearly seen for each measurement. Between 300 K and $\sim 170$ K, the  $c$-axis thermal expansion coefficient is about 3 times larger than that for the $a$-axis. Below the transition, between approximately 110 K and 80 K the $c$-axis thermal expansion coefficient is almost temperature independent. In is not clear if the apparent broad feature in $\alpha_c$ below $\sim 50$ K (Fig. \ref{F1}(b)) is an artifact of the measurements on a thin sample (i.e. a signal to noise problem), or is a real feature. This can be clarified by measurements on a significantly thicker samples. Our $a$-axis, high temperature ($T > 150$ K), thermal expansion data are consistent with the average $a$-axis thermal expansion coefficient in the same temperature range that can be evaluated from the x-ray data in \cite{rot09a} as well as with the relative changes of the $c$-lattice parameter between two temperature points reported in \cite{rot08b,hua08a}.

\begin{figure}
\begin{center}
\begin{minipage}{130mm}
\subfigure[]{
\resizebox*{6.5cm}{!}{\includegraphics{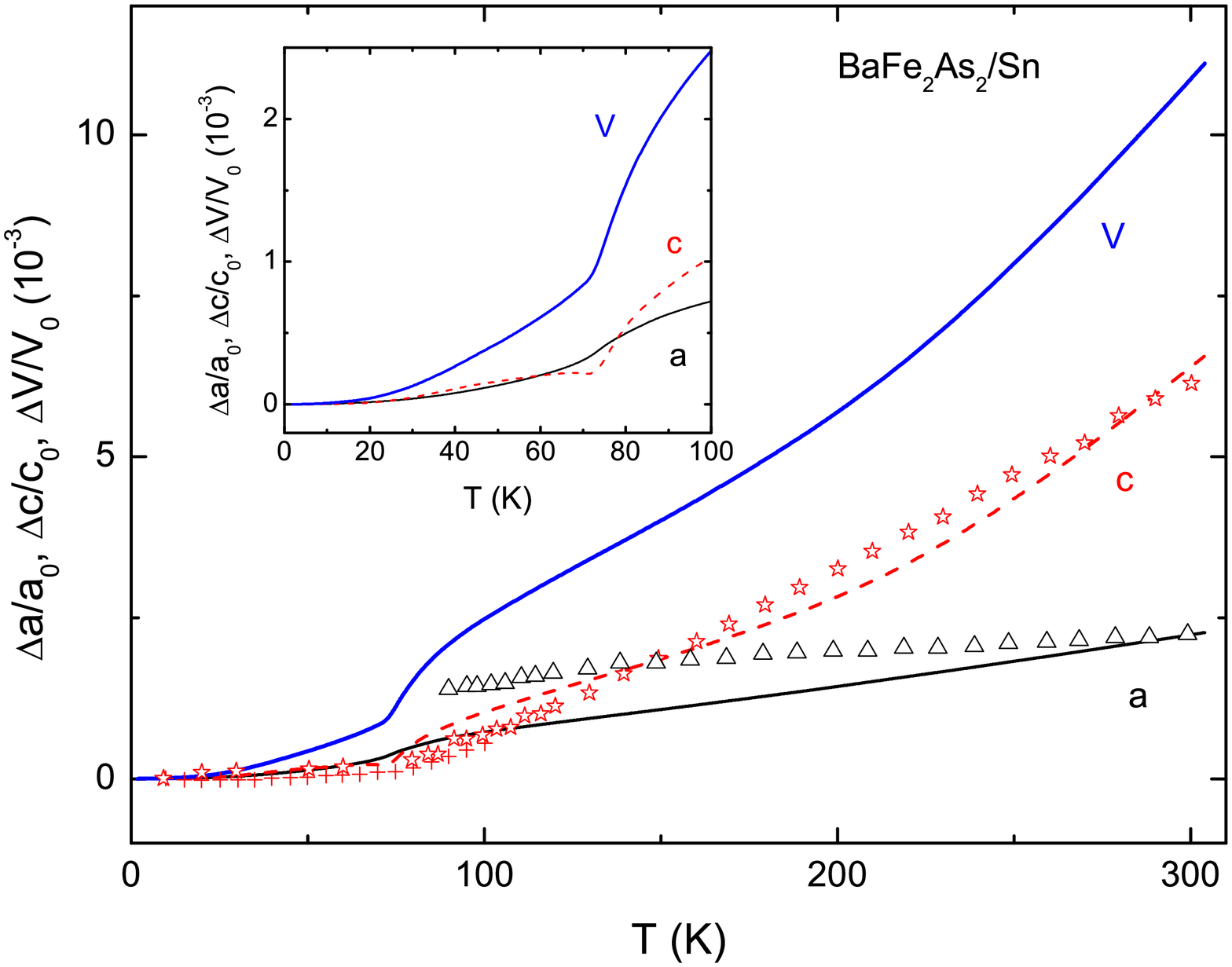}}}%
\subfigure[]{
\resizebox*{6.5cm}{!}{\includegraphics{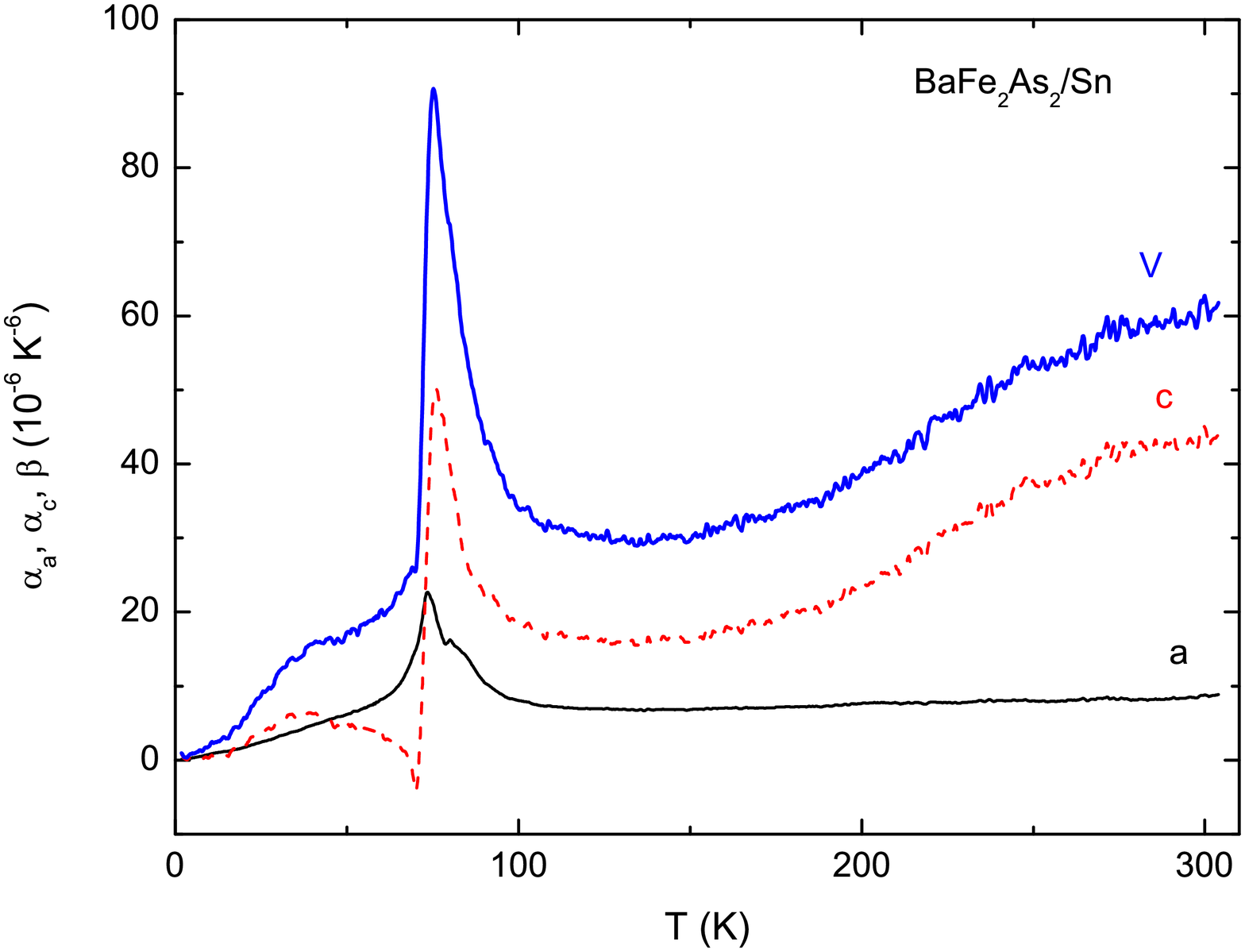}}}%
\caption{(a) Anisotropic thermal expansivities of BaFe$_2$As$_2$ grown out of Sn flux plotted relative to the values at $T = 1.8$ K; inset: enlarged low temperature part. Symbols: $a$-axis data from \cite{suy09a} (triangles, normalized at 300 K) and $c$-axis data from \cite{suy09a} (stars, normalized at $\sim 10$ K) and \cite{nin08b} (crosses, normalized at $\sim 10$ K) (b) Anisotropic thermal expansion coefficients of BaFe$_2$As$_2$ grown out of Sn flux.}%
\label{F2}
\end{minipage}
\end{center}
\end{figure}

BaFe$_2$As$_2$ crystals grown out of Sn flux have small amount of Sn from the flux incorporated into the structure. Although the exact site and amount of the Sn incorporation are still under debate, the physical properties reported by different groups are consistent. \cite{nin08b,suy09a,bae08a} As a result of Sn incorporation the structural/antiferromagnetic phase transition shifts down to $\sim 85 - 90$ K and becomes broader. This is also seen in the thermal expansion measurements (Fig. \ref{F2}). The features at the transition in anisotropic thermal expansivities are smoother and smaller. Thermal expansion coefficients above the transition, near room temperature, have values and anisotropy close to that observed in pure BaFe$_2$As$_2$ (Fig. \ref{F1}). The average high temperature (between 300 K and 150 K) thermal expansion coefficients can be also estimated from the published scattering data \cite{suy09a}. For the $c$-axis, both sets of data are very close, whereas for the $a$-axis the scattering data give the value of the average thermal expansion coefficient factor of 2 - 3 times lower than that presented here. This difference is probably extrinsic, related to the accuracies of the techniques, however the differences in the samples cannot be excluded.

\begin{figure}
\begin{center}
\begin{minipage}{130mm}
\subfigure[]{
\resizebox*{6.5cm}{!}{\includegraphics{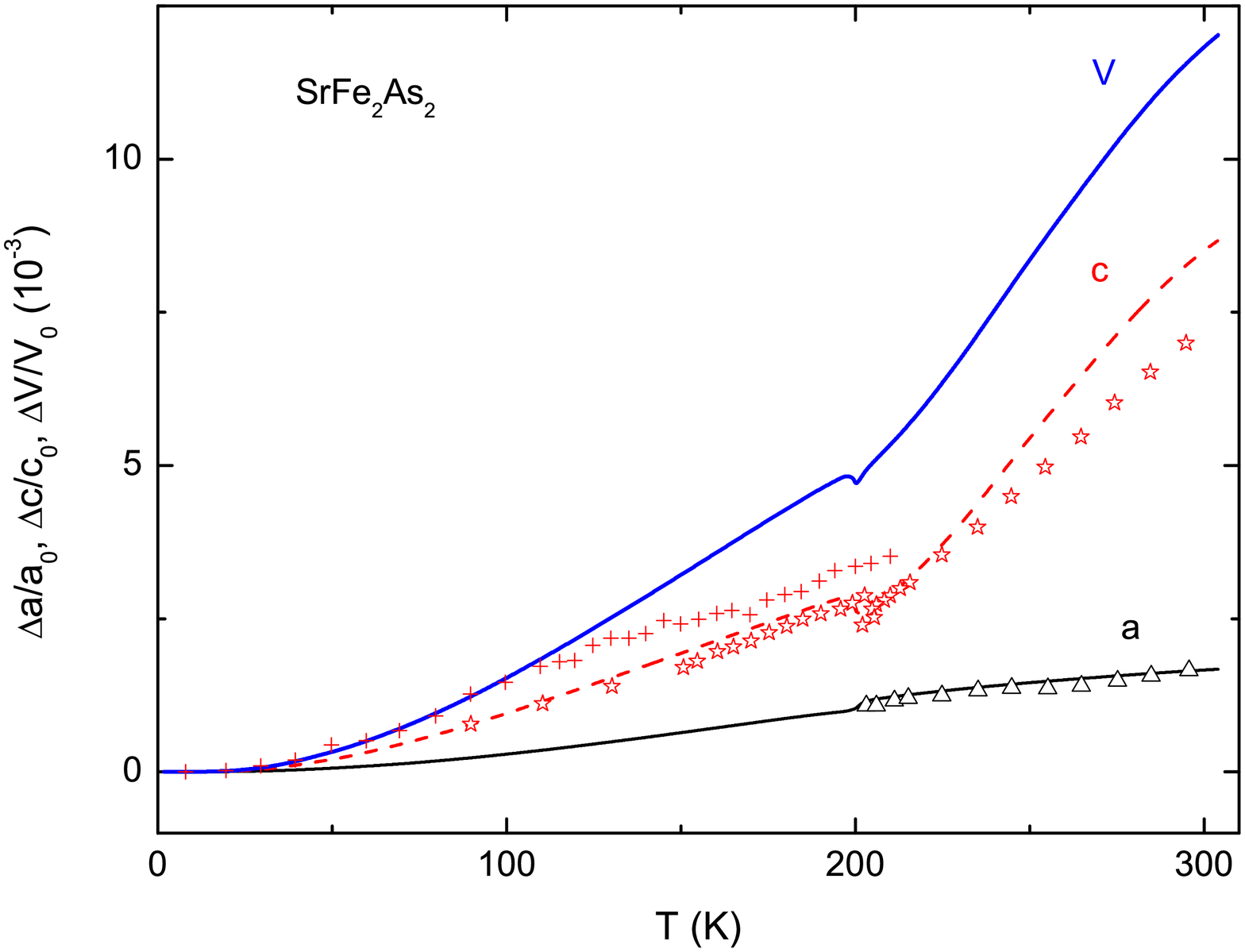}}}%
\subfigure[]{
\resizebox*{6.5cm}{!}{\includegraphics{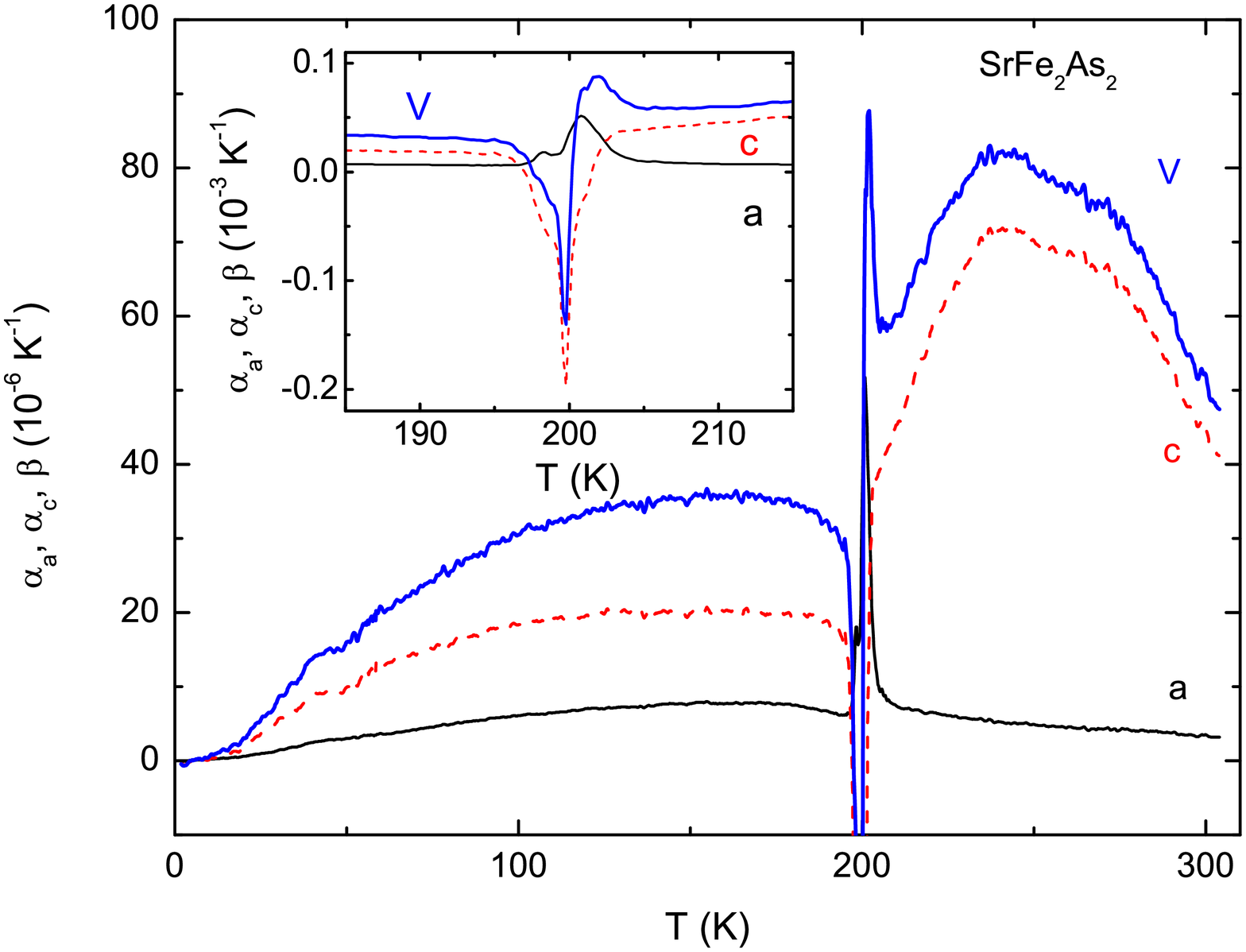}}}%
\caption{(a) Anisotropic thermal expansivities of SrFe$_2$As$_2$ plotted relative to the values at $T = 1.8$ K. Symbols: $a$-axis data from \cite{teg08a} (triangles, normalized at 296 K) and $c$-axis data from \cite{teg08a} (stars, normalized at $\sim 90$ K) and \cite{yan08a} (crosses, normalized at $\sim 8$ K) (b) Anisotropic thermal expansion coefficients of SrFe$_2$As$_2$; inset: enlarged region near structiral/magnetic phase transitions.}%
\label{F3}
\end{minipage}
\end{center}
\end{figure}

Data for SrFe$_2$As$_2$ are presented in Fig. \ref{F3}. The relative changes in the $c$-axis and in-plane dimensions on warming through the structural/antiferromagnetic transition are relatively small: the in-plane dimension increase is  $\sim 0.2 \cdot 10^{-3}$, whereas the $c$-axis decrease is $\sim 0.36 \cdot 10^{-3}$. The thermal expansion coefficients at temperatures above the structural/magnetic transition are even more anisotropic than for BaFe$_2$As$_2$ ($\alpha_a$ being somewhat smaller and $\alpha_c$ notably larger than those of BaFe$_2$As$_2$). Below the transition, in contrast to the BaFe$_2$As$_2$ data, $\alpha_c$ is larger that the in-plane thermal expansion coefficient. Our data are consistent with the average thermal expansion coefficients that one can evaluate from the scattering data \cite{yan08a,teg08a}.

\begin{figure}
\begin{center}
\begin{minipage}{130mm}
\subfigure[]{
\resizebox*{6.5cm}{!}{\includegraphics{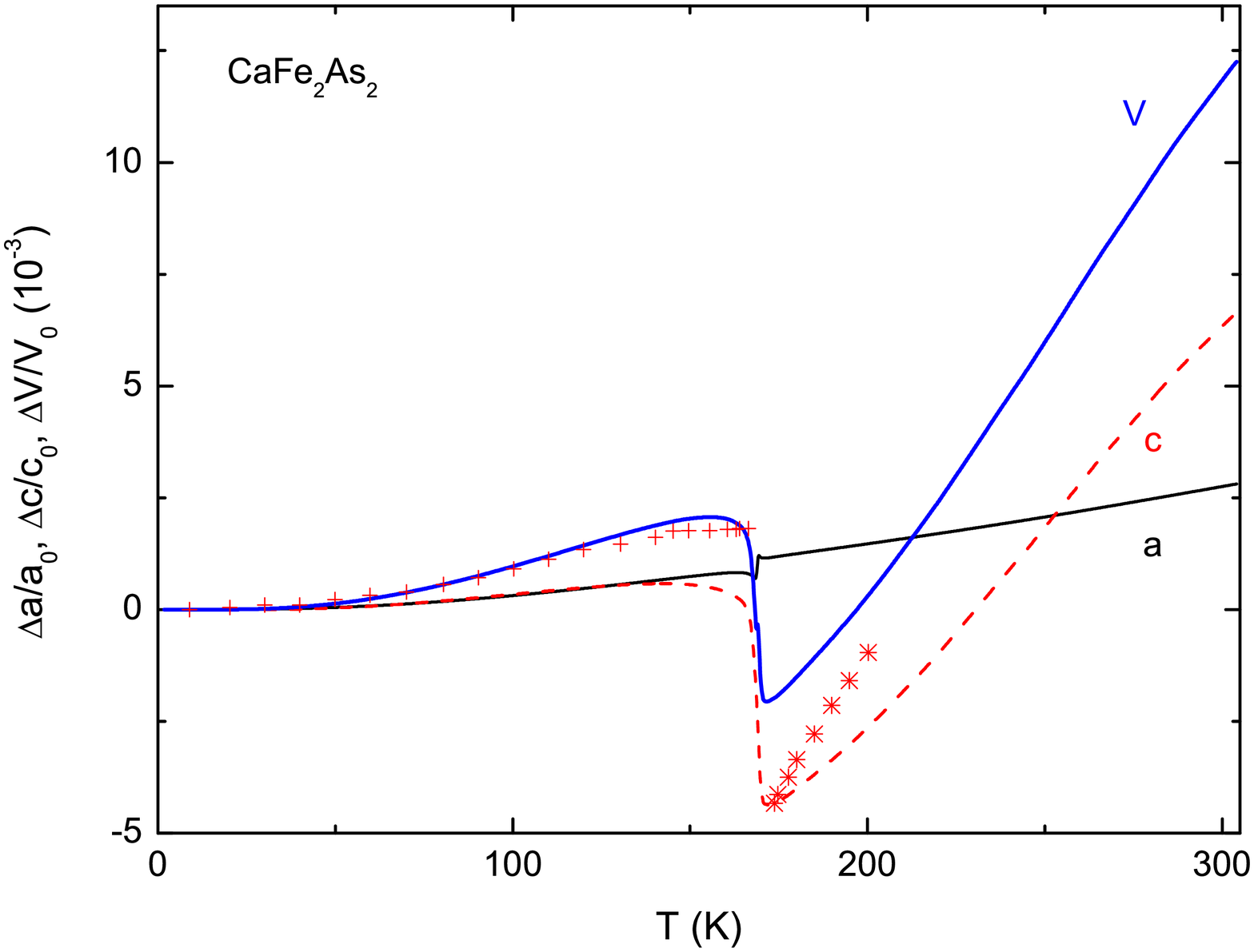}}}%
\subfigure[]{
\resizebox*{6.5cm}{!}{\includegraphics{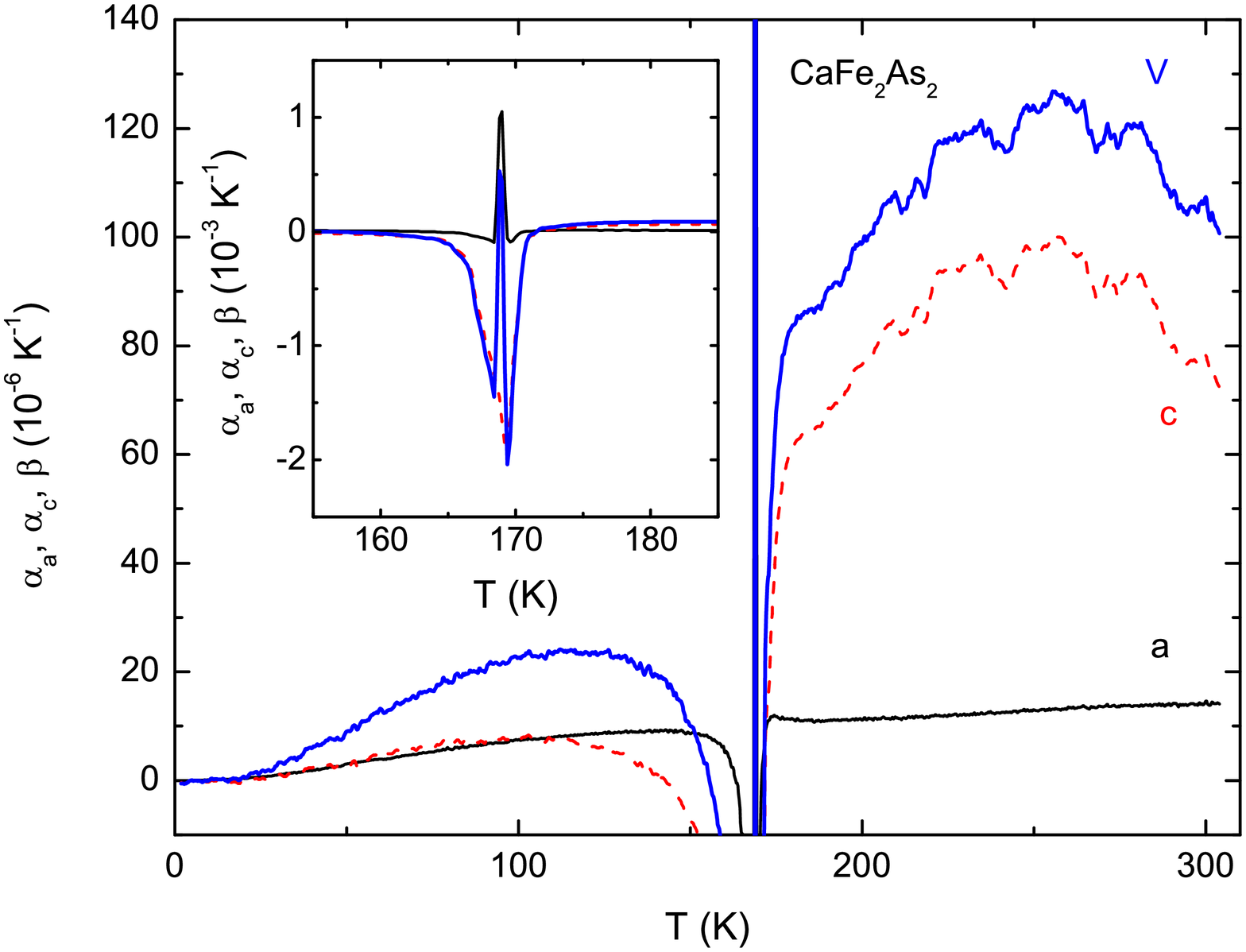}}}%
\caption{(a) Anisotropic thermal expansivities of CaFe$_2$As$_2$ plotted relative to the values at $T = 1.8$ K. Symbols: plotted separately (see the text), $c$-axis data below the structural/antiferromagnetic phase transition (crosses, normalized at $\sim 9$ K) and above the transition (asterisks, normalized at $\sim 174$ K) \cite{nin08a} (b) Anisotropic thermal expansion coefficients of CaFe$_2$As$_2$; inset: enlarged region near structiral/magnetic phase transitions.}%
\label{F4}
\end{minipage}
\end{center}
\end{figure}

The structural/antiferromagnetic phase transition in CaFe$_2$As$_2$ is clearly seen in thermal expansion data (Fig. \ref{F4}). The relative change of the $c$-axis on warming through the transition is very large, close to 0.5\%, whereas the measured change of the in-plane dimension is significantly smaller. The thermal expansion coefficients in the high temperature, tetragonal, phase are anisotropic and similar to those of SrFe$_2$As$_2$, and they appear to be almost isotropic below $\sim 100$ K. These coefficients are grossly consistent with the scattering data above and below the transition, \cite{nin08a}  but it should be noted that, in Fig. \ref{F4}a, the low temperature and high temperature scattering data are shown separately, normalized at $\sim 9$ K and $\sim 174$ K, respectively. In both temperature regions the $c$-axis thermal expansion coefficients inferred from the scattering data are larger than our data. The scattering data \cite{nin08a} show that both lattice parameters are larger in the high temperature, tetragonal phase ($a_{HT} > a_{LT} > b_{LT}$, $c_{HT} > c_{LT}$), that is in variance to our results for the $c$-axis close to the structural/antiferromagnetic transition. This striking difference in the observed change in the $c$-lattice parameter through the structural/antiferromagnetic transition can be, at least partially explained by the difficulties to unambiguously de-convolute the "real" change of the $c$-lattice parameter and effect of changes of crystal alignment/mosaicity on cooling/warming through such a dramatic, first order, structural phase transition in scattering experiments \cite{nin08a}. The assumptions used in the interpretation of the scattering data were outlined in the text and related figure captions of the Ref. \cite{nin08a}.
\\

In summary, the anisotropic thermal expansion data of the parent, {\it AE}Fe$_2$As$_2$ ({\it AE} = Ba, Sr, and Ca), compounds were measured. Above the structural/antiferromagnetic transition a significant anisotropy of the thermal expansion coefficients was observed, with $\alpha_a$ approximately 3-4 times smaller than $\alpha_c$, and the coefficients themselves have similar values for the parent compounds studied. The sharp anomalies associated with the structural/antiferromagnetic phase transitions are clearly seen in the thermal expansion measurements. For all three pure compounds the "average" $a$-value increases and the $c$-lattice parameter decreases on warming through the transition with the smallest change in the lattice parameters observed for SrFe$_2$As$_2$. One of the implications of these step-like changes in linear dimensions/lattice parameters on crossing the phase line is that non-negligible nonhydrostatic pressure components are expected at low temperatures when these materials are cooled down in a liquid media pressure cells (for e.g. CaFe$_2$As$_2$ see discussion in Refs. \cite{yuw09a,can09a}) with SrFe$_2$As$_2$ being the most tolerant to these conditions.  For the Sn-flux-grown BaFe$_2$As$_2$ the structural/antiferromagnetic phase transition shifts down in temperature and the features become significantly broader (similar broadening was observed for lightly Co-doped BaFe$_2$As$_2$ \cite{bud09a}).

Given the agreement with X-ray and neutron diffraction measurements, combined with the higher resolution and relative ease of the temperature cycling, thermal expansion measurements should be of great utility for tracking changes across {\it TM}-doping series and shedding the light on the physics underlining the $T - x$ phase diagrams \cite{can09b,nin09a}.

\section*{Acknowledgments}

Work at the Ames Laboratory was supported by the US Department of Energy - Basic Energy Sciences under Contract No. DE-AC02-07CH11358. We are indebted to George M. Schmiedeshoff for his help in establishing dilatometry technique in Ames Laboratory Novel Materials and Ground States Group and for many propitious advices. We thank Andreas Kreyssig  and Shibabrata Nandi  for useful discussions and Jiaqiang Yan for help in synthesis.

\medskip

\end{document}